\documentclass[11pt,a4paper]{article}

\usepackage{jheppub}
\usepackage{graphicx}
\usepackage{dcolumn}
\usepackage{bm}
\usepackage{amsmath}
\usepackage{braket}
\usepackage{slashed}
\usepackage{epstopdf}
\usepackage{placeins}
\usepackage{multirow}
\usepackage{makecell}
\usepackage{soul}
\usepackage{braket}
\usepackage[shortlabels]{enumitem}
\usepackage{placeins}

\usepackage[labelfont=bf, textfont=it, font=small]{caption}
\newcommand{\beq}{\begin{eqnarray}}
\newcommand{\eeq}{\end{eqnarray}}
\newcommand{\beqnn}{\begin{eqnarray*}}
\newcommand{\eeqnn}{\end{eqnarray*}}

\renewcommand{\thesection}{\arabic{section}}

\newcommand{\Tr}{\ensuremath{\mathrm{Tr}}}

\newcommand{\YM}{\ensuremath{\mathrm{YM}}}

\newcommand{\cool}{\ensuremath{\mathrm{cool}}}

\newcommand{\SU}{\ensuremath{\mathrm{SU}}}

\newcommand{\QCD}{\ensuremath{\mathrm{QCD}}}

\newcommand{\clov}{\ensuremath{\mathrm{clov}}}

\begin{document}
	
\title{The topological susceptibility slope $\chi^\prime$ of the pure-gauge SU(3) Yang--Mills theory}

\author[a]{Claudio Bonanno}

\affiliation[a]{Instituto de F\'isica Te\'orica UAM-CSIC, c/ Nicol\'as Cabrera 13-15, Universidad Aut\'onoma de Madrid, Cantoblanco, E-28049 Madrid, Spain}

\emailAdd{claudio.bonanno@csic.es}

\abstract{We determine the pure-gauge $\mathrm{SU}(3)$ topological susceptibility slope $\chi^\prime$, related to the next-to-leading-order term of the momentum expansion of the topological charge density 2-point correlator, from numerical lattice Monte Carlo simulations. Our strategy consists in performing a double-limit extrapolation: first we take the continuum limit at fixed smoothing radius, then we take the zero-smoothing-radius limit. Our final result is $\chi^\prime = [17.1(2.1)~\mathrm{MeV}]^2$. We also discuss a theoretical argument to predict its value in the large-$N$ limit, which turns out to be remarkably close to the obtained $N=3$ lattice result.}


\maketitle

\flushbottom

\section{Introduction}\label{sec:intro}

Topological properties are among the most interesting non-perturbative features of $4d$ $\SU(N)$ gauge theories, both for their theoretical implications and for their relation with phenomenological aspects of strong interactions. In this respect, an interesting topological quantity to investigate is the Euclidean topological charge density 2-point correlator~\cite{Alles:1997ae,Vicari:1999xx,Horvath:2005cv,Vicari:2008jw,Chowdhury:2012sq,Fukaya:2015ara,Mazur:2020hvt,Altenkort:2020axj,BarrosoMancha:2022mbj}:
\beq\label{eq:corr_def}
\widetilde{G}(p^2) \equiv \int d^4 x\, e^{i p \cdot x} \braket{q(x)q(0)}, \qquad \qquad p^2 \equiv p_\mu p_\mu,
\eeq
where $q(x)$ is the $\SU(N)$ topological charge density,
\beq\label{eq:topcharge_dens_def}
q(x) = \frac{1}{16 \pi^2} \Tr\left\{F_{\mu\nu}(x)\widetilde{F}_{\mu\nu}(x)\right\}.
\eeq

In order to better discuss the role played by this quantity, let us first introduce the momentum-expansion of~\eqref{eq:corr_def} in powers of $p^2$:
\beq\label{eq:corr_expansion}
\widetilde{G}(p^2) = \widetilde{G}(0) + \frac{\widetilde{G}(p^2)}{dp^2}\Bigg\vert_{p^2\,=\,0} p^2 + \mathcal{O}(p^4).
\eeq
The leading-order term $\widetilde{G}(0)$ is the well-known \emph{topological susceptibility} $\chi$:
\beq\label{eq:chi_def}
\widetilde{G}(0) = \int d^4 x\, \braket{q(x)q(0)} = \lim_{V\to\infty} \frac{\braket{Q^2}}{V} \equiv \chi, \qquad \qquad Q = \int d^4 x \, q(x),
\eeq
while the next-to-leading-order term, which is the main object of study of this paper, is the so-called \emph{topological susceptibility slope} $\chi^\prime$:
\beq\label{eq:chiprime_def}
\chi^\prime \equiv - \frac{\widetilde{G}(p^2)}{dp^2}\Bigg\vert_{p^2\,=\,0} = \frac{1}{8} \int d^4 x \, x^2 \braket{q(x)q(0)}, \qquad \qquad x^2 \equiv x_\mu x_\mu.
\eeq

As far as theoretical and phenomenological physical aspects of QCD are concerned, the topological susceptibility slope is an extremely interesting quantity to investigate. For example, in the large-$N$ limit, the value of $\chi^\prime$ controls the internal consistency of the Witten--Veneziano mechanism. Indeed, for the Witten--Veneziano equation to hold, the susceptibility slope has to satisfy the following condition: $\vert \chi^\prime \vert m^2_{\eta^\prime} \ll \chi$~\cite{Witten:1978bc,Veneziano:1979ec,DiVecchia:1980yfw}. In full QCD, instead, it is possible to relate the chiral limit of $\chi^\prime$ to the nucleon axial charge, an experimentally-measurable hadronic quantity, via the $\mathrm{U}(1)$ Goldberger--Treiman relation~\cite{Shore:1990zu,Shore:1992rg,Alles:1995aw,Narison:1998aq,Bernard:2001rs}.

Analytical predictions of the value of $\chi^\prime$ can be obtained using suitable approximations. For example, results for this quantity have been obtained using Chiral Perturbation Theory~\cite{Leutwyler:2000jg} and the QCD Sum Rule~\cite{Ioffe:1998sa,Narison:1998aq,Narison:2006ws,Narison:2021svo}, which is also able to provide an estimate for the pure-gauge $\SU(3)$ Yang--Mills theory~\cite{Narison:1998aq,Narison:2006ws}. On the other hand, given the purely non-perturbative nature of the topological properties of gauge theories, a natural first principle to compute $\chi^\prime$ is provided by Monte Carlo (MC) simulations on the lattice. However, for what concerns the numerical studies of the susceptibility slope, the current state of the art is pretty different compared to other topological quantities like the topological susceptibility. As a matter of fact, while $\chi$ has been extensively studied from the lattice in full QCD~\cite{Bonati:2015vqz, Frison:2016vuc, Borsanyi:2016ksw, Petreczky:2016vrs, Bonati:2018blm,Burger:2018fvb,Bonanno:2019xhg,Lombardo:2020bvn, Kotov:2021ujj, Athenodorou:2022aay, Chen:2022fid}, in $\SU(N)$ pure Yang--Mills theories~\cite{Alles:1996nm, Alles:1997qe, DelDebbio:2004ns, DelDebbio:2002xa, DElia:2003zne,DelDebbio:2006yuf, Lucini:2004yh,Giusti:2007tu, Vicari:2008jw,Luscher:2010ik, Panagopoulos:2011rb, Ce:2015qha, Ce:2016awn,Bonati:2015sqt, Bonati:2016tvi, Bonati:2018rfg, Burger:2018fvb,Bonati:2019kmf,Athenodorou:2020ani,Bonanno:2020hht,Athenodorou:2021qvs,Bonanno:2022yjr,Bennett:2022ftz}, and in other lower-dimensional models~\cite{Campostrini:1988cy,Campostrini:1992ar,Campostrini:1992it,Alles:1997nu,DelDebbio:2004xh,Bietenholz:2010xg,Hasenbusch:2017unr,Bonati:2017woi,Bonanno:2018xtd,Berni:2019bch,Berni:2020ebn,Bonanno:2022dru}, only few preliminary numerical lattice studies of the topological susceptibility slope in QCD and in the pure-gauge $\SU(3)$ can be retrieved in the literature~\cite{DiGiacomo:1990ij,Briganti:1991pb,Digiacomo:1992jg,Boyd:1997nt,Koma:2010vx}.

Given the interesting theoretical and phenomenological implications of $\chi^\prime$, the goal of this work is to go beyond the state of the art earlier outlined, and provide a solid numerical determination of the topological susceptibility slope $\chi^\prime$ in the $\SU(3)$ pure-gauge theory from lattice MC simulations. To do so, we will follow the exact same strategy that we applied in~\cite{Bonanno:2022hmz}, where we provided lattice determinations of $\chi^\prime$ in $2d$ $\mathrm{CP}^{N-1}$ models for a wide range of $N$ values, which were shown to be in excellent agreement with analytic predictions obtained from the large-$N$ $1/N$ expansion.

Let us briefly outline the numerical procedure we will follow. First, we determine $\chi^\prime$ at finite lattice spacing from Eq.~\eqref{eq:chiprime_def} on smoothened lattice gauge configurations. Smoothing is a necessary procedure to remove ultraviolet (UV) noise and correctly identify the physical topological background of gauge configurations. On the other hand, however, it modifies the short-distance behavior of the correlators, introducing a non-trivial dependence of $\chi^\prime$ on the amount of smoothing performed. Thus, to recover the correct physical value of $\chi^\prime$, we have to perform a double extrapolation: first, we take the continuum limit of $\chi^\prime$ at fixed smoothing radius; then, we take the zero-smoothing-radius limit, which eventually yields the physical result for the susceptibility slope. The second limit is needed to ensure that no relevant physical contribution to $\chi^\prime$ from short distances is lost.

This paper is organized as follows: in Sec.~\ref{sec:setup} we briefly summarize our lattice setup and we describe in details our double-limit strategy; in Sec.~\ref{sec:results} we present and discuss our numerical lattice calculation of $\chi^\prime$; finally, in Sec.~\ref{sec:conclu} we draw our conclusions and discuss possible future outlooks of this work.

\section{Lattice setup}\label{sec:setup}

In this section we will present our lattice discretization and, most importantly, the numerical strategy we pursued to compute $\chi^\prime$.

\subsection{Lattice action and parameters}\label{sec:lattice_discr}

We discretize the pure-gauge $\SU(3)$ action on a $L^4$ hyper-cubic lattice with periodic boundary conditions using the standard Wilson plaquette action:
\beq\label{eq:Wilson_action}
S_{\mathrm{W}}[U] \equiv - \frac{\beta}{3} \sum_{x,\mu>\nu} \Re\Tr\left\{U_\mu(x) U_\nu(x+a\hat{\mu}) U_\mu^\dagger(x+a\hat{\nu})U_\nu^\dagger(x)\right\},
\eeq
where $a$ is the lattice spacing, $\beta = 6/g^2$ is the bare coupling and $U_\mu(x)$ are the $\SU(3)$ gauge link variables. We performed MC simulations for 5 values of the bare coupling using a 4:1 mixture of over-relaxation~\cite{Creutz:1987xi} and over-heat-bath~\cite{Creutz:1980zw,Kennedy:1985nu} local updating sweeps implemented \emph{\'a la} Cabibbo--Marinari~\cite{Cabibbo:1982zn}, i.e., updating the 3 $\SU(2)$ subgroups of $\SU(3)$. In the following we will refer to this combination as our single MC updating step. The scale was set through the Sommer parameter $r_0$, and the lattice sizes were chosen to ensure that $aL/r_0 \sim 3$, i.e., $aL\sim 1.4$ fm, which we will show to be enough to contain finite-size effects within our typical statistical error. Simulation parameters and the total accumulated statistics are reported in Tab.~\ref{tab:simul_params}. As we will show in the following, topological freezing is not an issue for the explored range of $\beta$, thus no specific strategy to mitigate it was necessary.

\begin{table}[!htb]
\begin{center}
\begin{tabular}{|c|c|c|c|c|c|c|}
\hline
$\beta$ & $L$ & $a/r_0$ & $a$ [fm] & $aL/r_0$ & $aL$ [fm] & Stat. \\
\hline
5.95 & 16 & 0.2036(5) & 0.0961 & 3.26 & 1.54 & 1.94M \\
6.00 & 16 & 0.1863(5) & 0.0879 & 2.98 & 1.41 & 2.03M \\
6.07 & 18 & 0.1656(4) & 0.0782 & 2.98 & 1.41 & 1.94M \\
6.20 & 22 & 0.1354(5) & 0.0639 & 2.98 & 1.41 & 1.23M \\
6.40 & 30 & 0.1027(5) & 0.0485 & 3.07 & 1.45 & 0.71M \\
\hline
\end{tabular}
\end{center}
\caption{Summary of simulation parameters. Scale setting was performed according to the determination of $a(\beta)/r_0$ of Ref.~\cite{Necco:2001xg}. In order to pass from $r_0$ to physical units, we used the value of the Sommer scale $r_0 = 0.472(5)~\mathrm{fm}$ given in Ref.~\cite{Sommer:2014mea}. Measures of topological quantities were taken every 10 MC updating steps for all values of $\beta$ but the largest, for which measures were taken every 100 MC steps. Total collected statistics is expressed in millions (M).}
\label{tab:simul_params}
\end{table}

\subsection{The smoothing radius dependence of the susceptibility slope}\label{sec:lattice_topology}

In this work we will adopt the simplest discretization of the topological charge density~\eqref{eq:topcharge_dens_def} with definite parity, the so-called \emph{clover} discretization:
\beq
q_\clov(x) &=&  \frac{-1}{2^9 \pi^2}\sum_{\mu\nu\rho\sigma=\pm1}^{\pm4}\varepsilon_{\mu\nu\rho\sigma}
\Tr\left\{\Pi_{\mu\nu}(n)\Pi_{\rho\sigma}(n)\right\},\\
\nonumber\\
Q_\clov &=& \sum_{x} q_\clov(x),
\eeq
with $\Pi_{\mu\nu}(x)=U_\mu(x) U_\nu(x+\hat{\mu}) U_\mu^\dagger(x+\hat{\nu})U_\nu^\dagger(x)$ the plaquette and $\varepsilon_{\mu\nu\rho\sigma}$ the Levi--Civita symbol, satisfying $\varepsilon_{(-\mu) \nu \rho \sigma} =-\varepsilon_{\mu \nu \rho \sigma}$. This definition of the lattice topological charge possess the correct naive continuum limit but is not integer-valued at finite lattice spacing, as it can be shown to renormalize multiplicatively~\cite{Campostrini:1988cy,Vicari:2008jw}:
\beq
Q_\clov = Z_Q(\beta) Q.
\eeq
The related lattice topological susceptibility,
\beq
a^4\chi_\clov = \frac{\braket{Q_\clov^2}}{V}, \qquad V=L^4,
\eeq
instead, can be shown to renormalize both multiplicatively and additively~\cite{DiVecchia:1981aev,DElia:2003zne,Vicari:2008jw}:
\beq
\chi_\clov = Z_Q^2(\beta) \chi + M(\beta),
\eeq
where $M$ arises from short-distance contact terms~\cite{DiVecchia:1981aev,DElia:2003zne,Vicari:2008jw}. The additive term $M(\beta)$ diverges in the continuum limit, eventually disrupting the physical topological signal.

To get rid of renormalization effects, it is customary to compute lattice topological quantities on smoothened configurations. Smoothing algorithms are used to dampen UV fluctuations up to a length scale, known as the \emph{smoothing radius} $r_s$, which is proportionally to the square root of the amount of smoothing performed. Therefore, if not overly prolonged, smoothing is expected to leave the relevant long-distance topological fluctuations, i.e., the global topological content of a gauge configuration $Q$, untouched.

On smoothened configurations, thus, $Z\simeq 1$ and $M\simeq 0$.; therefore, defining:
\beq
q_L(x) &=& q_\clov^{(\mathrm{smooth})}(x),\\
&&\nonumber\\
Q_L &=& \sum_x q_L(x),\\
&&\nonumber\\
a^4 \chi_L &=& \frac{\braket{Q^2_L}}{V}, \qquad V=L^4,
\eeq
where $q_{\clov}^{(\mathrm{smooth})}(x)$ stands for $q_\clov(x)$ computed on smoothened gauge fields, the distribution of $Q_L$ will show sharp peaks close to integer values, and the lattice susceptibility $\chi_L$ will be free of short-distance singularities when approaching the continuum limit.

In the light of this discussion it is thus clear that, if we take the continuum limit of $\chi_L$ at fixed smoothing radius, the obtained result should be independent of the chosen smoothing radius, as long as it is taken sufficiently small that there is an effective separation between the UV scale of the smoothened fluctuations and the infrared (IR) scale of the relevant topological fluctuations.

The dependence on the smoothing radius of $\chi^\prime$ is instead different. As a matter of fact, now we are dealing with an observable which does not depend on the global topological charge $Q$, but rather on the integral of the topological charge density 2-point correlator, cf.~Eq.~\eqref{eq:chiprime_def}. Since smoothing has the unavoidable effect of modifying the short-distance behavior of the correlator, it is natural to expect a non-trivial dependence of the continuum-extrapolated results for $\chi^\prime$ on the smoothing radius. Clearly, the physical value of the susceptibility slope sits at $r_s=0$, where no contribution from short distances is smoothened away. Thus, in this case, in order to recover the proper physical value of $\chi^\prime$, one has also to take the zero-smoothing-radius limit after the continuum one.

In practice, if we define our lattice discretization of $\chi^\prime$ as,
\beq\label{eq:chip_lat}
a^2\chi^\prime_L \equiv \frac{1}{8} \left\langle\sum_x d^2(x,0) q_L(x) q_L(0)\right\rangle,
\eeq
with $d^2(x,y)$ the physical squared distance between lattice sites $x$ and $y$,
\begin{equation}
\begin{gathered}
d^2(x,y) = \sum_{\mu} d_\mu^2(x,y),\\
d^2_\mu(x,y) = 
\begin{cases}
(x_\mu - y_\mu)^2         ,& \vert x_\mu - y_\mu \vert \leq L/2, \\
[ L - (x_\mu - y_\mu) ]^2 ,& \vert x_\mu - y_\mu \vert > L/2,
\end{cases}
\end{gathered}
\end{equation}
we expect the continuum limit of $\chi^\prime_L(a,r_s)$, taken at fixed smoothing radius, to depend on the choice of $r_s$. Such dependence can be easily derived from the results of Ref.~\cite{Altenkort:2020axj}, where it was shown that, in the continuum theory, the finite-smoothing-radius 2-point correlator of the topological charge density is affected by leading $O(r_s^2)$ corrections compared to the zero-smoothing one. Thus, once the continuum limit is taken, we expect:
\beq\label{eq:zerosmooth_limit}
\chi^\prime(r_s) = \chi^\prime + k\, r_s^2 + o(r_s^2),
\eeq
where here $\chi^\prime$ denotes the zero-smoothing limit of $\chi^\prime(r_s)$, i.e., the actual physical value of the topological susceptibility slope.

So far, we have kept our discussion general, now we want to actually specify the particular smoothing method we are going to employ in the following. In this paper we will adopt \emph{cooling}~\cite{Berg:1981nw,Iwasaki:1983bv,Itoh:1984pr,Teper:1985rb,Ilgenfritz:1985dz,Campostrini:1989dh,Alles:2000sc} for its simplicity and numerical cheapness. This smoothing method has been shown to be numerically equivalent to other smoothing algorithms~\cite{Alles:2000sc, Bonati:2014tqa, Alexandrou:2015yba}, such as gradient flow~\cite{Luscher:2009eq, Luscher:2010iy} or stout smearing~\cite{APE:1987ehd, Morningstar:2003gk}. One cooling steps consists in performing a lattice sweep where each gauge link is aligned to its related local staple, so that the Wilson action~\eqref{eq:Wilson_action} is locally minimized. For the Wilson plaquette action, the relation between the number of cooling steps $n_\cool$ and the smoothing radius in lattice units is given by~\cite{Bonati:2014tqa}:
\beq\label{eq:smooth_radius}
\frac{r_s}{a} = \sqrt{\frac{8}{3} n_\cool}.
\eeq
It is thus clear that, according to Eq.~\eqref{eq:smooth_radius}, we expect our continuum results for $\chi^\prime$ to exhibit a linear dependence on $n_\cool$ (once it is fixed in physical units), cf.~Eq.~\eqref{eq:zerosmooth_limit}\footnote{See also Refs.~\cite{Bonanno:2022dru,Bonanno:2022hmz,Bonanno:2023ljc,Bonanno:2023thi} for other works where a linear extrapolation in the number of cooling steps (in physical units) is employed to compute a variety of topological observables.}.

In order to fix the smoothing radius in physical units, the number of cooling steps has to be scaled as $a^2$ among ensembles generated at different values of the bare coupling $\beta$, i.e., to fix $r_s$ it is sufficient to keep $n_\cool \times (a/r_0)^2 = a^2 n_\cool /r_0^2 \propto (r_s/r_0)^2$ constant among ensembles generated at different values of $\beta$.

\section{Results}\label{sec:results}

In this section we will discuss our numerical results. In the first part, we will present our lattice determination of the topological susceptibility slope by discussing our double-limit extrapolation of $\chi^\prime_L$, and by showing that several sources of systematical errors are well under control. In the second part, we will extensively discuss our result for $\chi^\prime$, comparing it with available analytic estimations, and presenting a theoretical argument to predict its value and scaling in the large-$N$ limit.

\subsection{The double limit extrapolation of $\chi^\prime$}

\begin{table}[!htb]
\begin{center}
\begin{tabular}{|c|c|c|c|c|c|c|}
\hline
$\beta$ & $L$ & $a/r_0$ & $n_\cool$ max & $\Delta n_\cool$ & $a^2 n_\cool/r_0^2 $ max & $\Delta [a^2 n_\cool/r_0^2]$ \\
\hline
5.95 & 16 & 0.2042(5) & 19 & 1 & 0.792 & 0.041 \\
6.00 & 16 & 0.1863(5) & 22 & 2 & 0.763 & 0.069 \\
6.07 & 18 & 0.1658(4) & 28 & 2 & 0.769 & 0.055 \\
6.20 & 22 & 0.1355(5) & 42 & 3 & 0.771 & 0.055 \\
6.40 & 30 & 0.1027(5) & 75 & 5 & 0.791 & 0.053 \\
\hline
\end{tabular}
\end{center}
\caption{Employed ranges of cooling steps to compute the topological susceptibility slope on the lattice. Measures of topological quantities where taken every $\Delta n_\cool$ cooling steps.}
\label{tab:cool_steps}
\end{table}

\begin{figure}[!htb]
\centering
\includegraphics[scale=0.45]{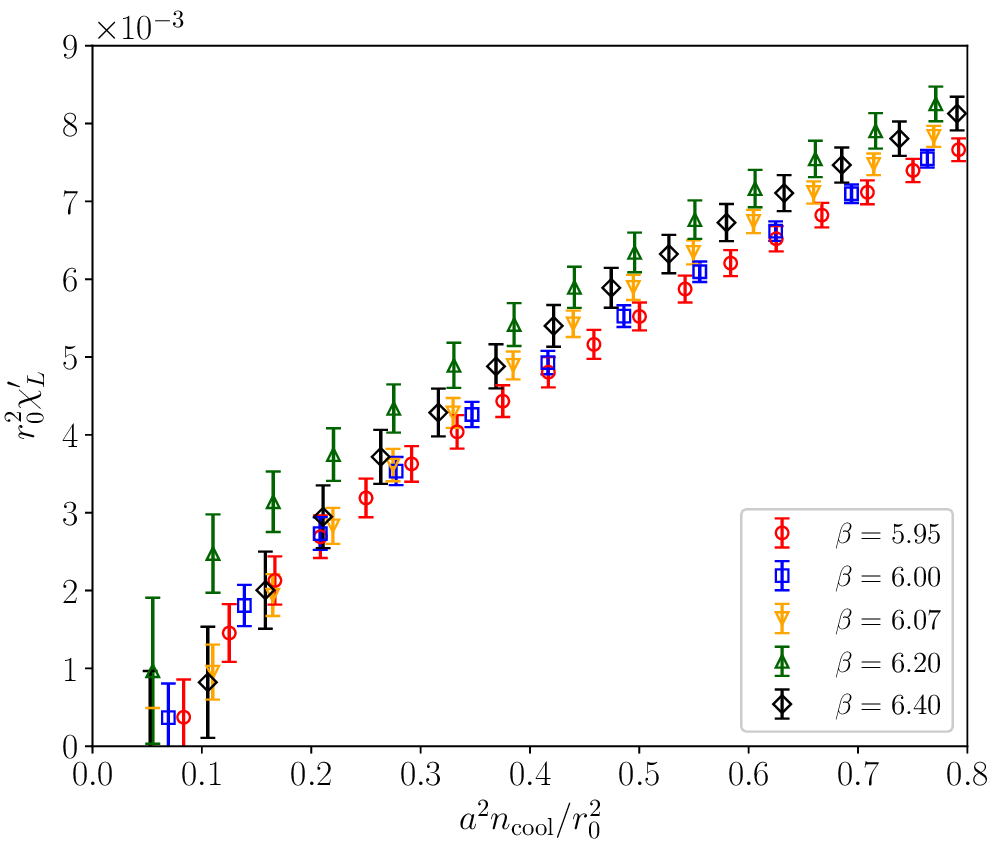}
\caption{Determinations of $r_0^2 \chi_L^\prime$ as a function of the smoothing radius, reported in terms of $a^2 n_\cool/r_0^2$, for all explored values of $\beta$.}
\label{fig:comp_chip}
\end{figure}

We compute the lattice topological susceptibility slope $r_0^2 \chi^\prime_L$ using the definition in Eq.~\eqref{eq:chip_lat} for several number of cooling steps, chosen in order to roughly correspond to the same range of physical smoothing radii, which in the following we will always express in terms of $a^2 n_\cool/ r_0^2$. In Tab.~\ref{tab:cool_steps} we report the ranges of cooling steps explored, while in Fig.~\ref{fig:comp_chip} we plot and compare our determinations obtained for different values of $\beta$ and $n_\cool$.

\begin{figure}[!t]
\centering
\includegraphics[scale=0.38]{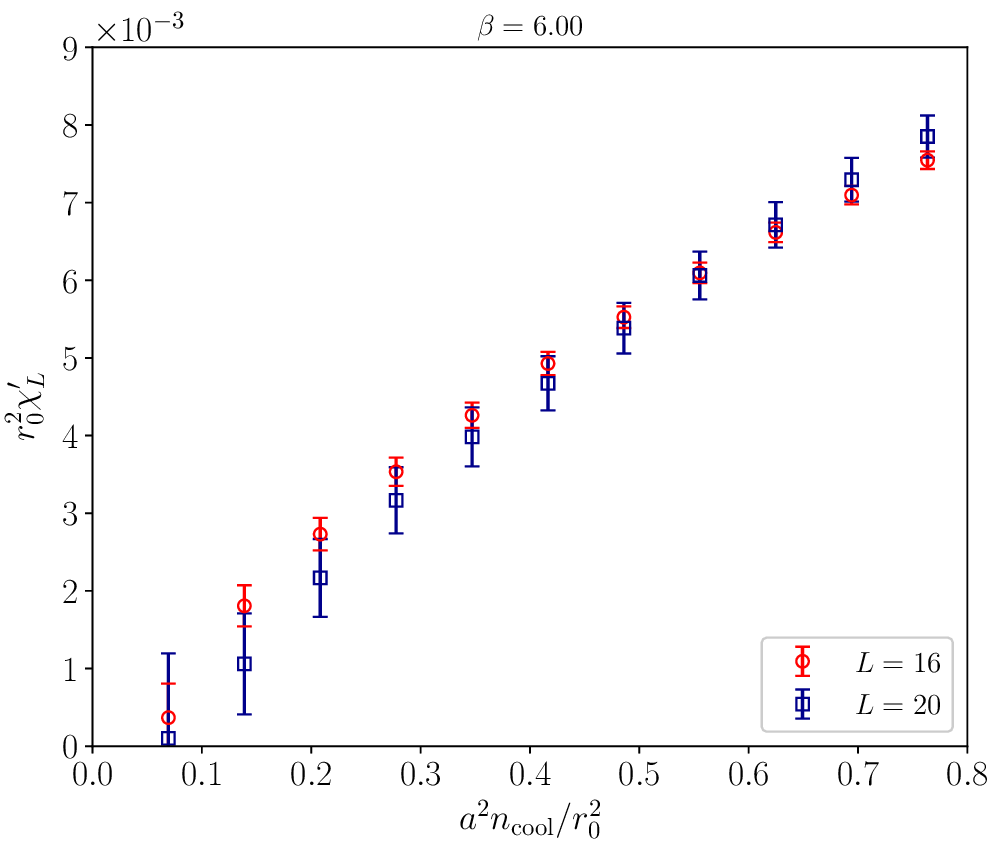}
\caption{Determinations of $r_0^2 \chi_L^\prime$ as a function of $a^2 n_\cool / r_0^2$ for $\beta=6.00$ and for 2 values of the lattice size, $L=16$ and $L=20$, corresponding respectively to $aL/r_0 \simeq 2.98$ and $3.73$. The statistics of the $L=20$ run is $\sim 1.94$M measures, i.e., roughly the same of the $L=16$ one.}
\label{fig:fse_chip}
\end{figure}

In order to verify that possible systematic effects coming from the choice of the lattice size are under control, we performed, for $\beta=6.00$, an additional calculation on a larger volume, obtaining perfectly agreeing results. Determinations of $r_0^2 \chi_L$ obtained for $L=16$ and $L=20$ are shown in Fig.~\ref{fig:fse_chip}. Thus, $r_0L \gtrsim 3$ is large enough to contain finite size effects within our typical statistical errors in the explored range of smoothing radii.

\begin{figure}[!t]
\centering
\includegraphics[scale=0.38]{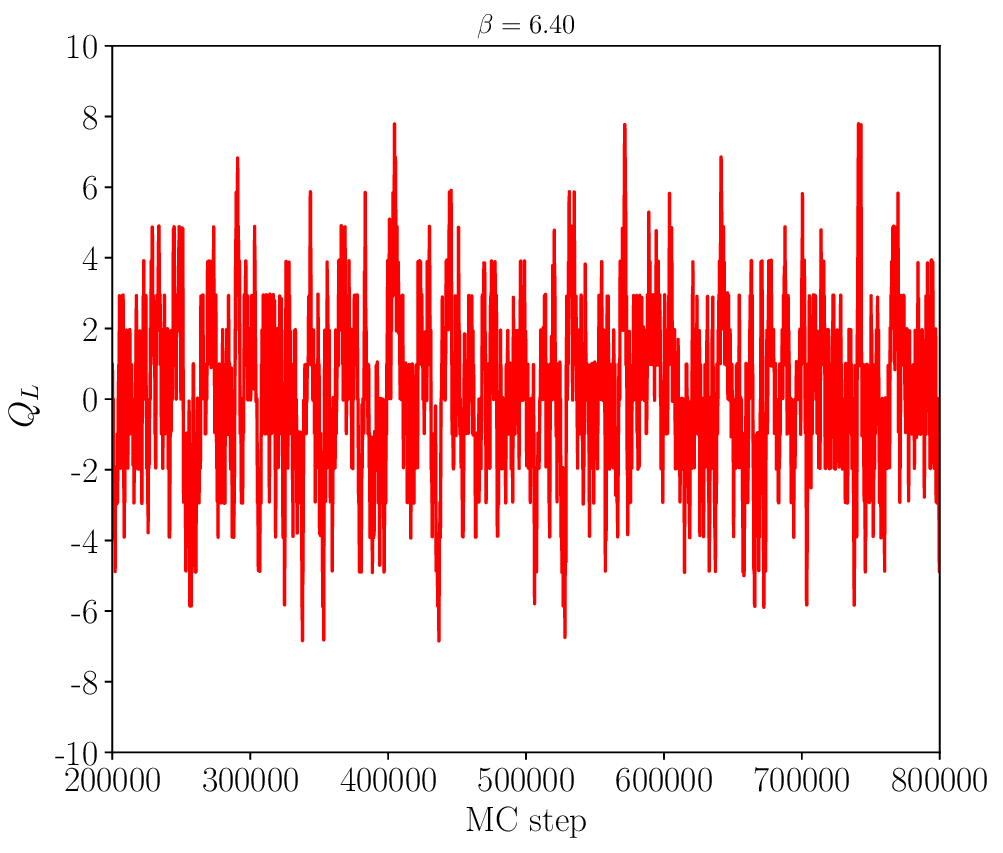}
\caption{History of the topological charge $Q_L$, obtained for $n_\cool=40$, for the finest lattice spacing explored in this work ($\beta=6.40$). The horizontal axis is expressed in units of the standard MC step defined in Sec.~\ref{sec:lattice_discr}, and corresponds to the $\sim 0.86\%$ of the total collected statistics.}
\label{fig:Q_history}
\end{figure}

Another possible source of systematic errors is topological freezing. Indeed, although $\chi^\prime_L$ could in principle be computed even without any fluctuation of the global topological charge, in~\cite{Bonanno:2022hmz} it has been shown that calculating the topological susceptibility slope within just the $Q=0$ sector would suffer from much larger finite size effects. As it can be observed from the plot in Fig.~\ref{fig:Q_history}, where a small fraction of the MC history of the topological charge $Q_L$ for our finest lattice spacing ($\beta=6.40$) is depicted, we observe lots of fluctuations of $Q_L$ during our MC evolution, so topological freezing is never an issue for the explored range of $\beta$.

We now move to discuss the continuum limit of $r_0^2 \chi^\prime_L$ at fixed smoothing radius. We extrapolated our data towards the continuum limit by performing a best fit of our data according to the following fit function:
\beq
r_0^2 \chi^\prime_L\left(a, \frac{a^2 n_\cool}{r_0^2}\right) = r_0^2 \chi^\prime\left(\frac{a^2 n_\cool}{r_0^2}\right) + c\left(\frac{a^2 n_\cool}{r_0^2}\right)\, \frac{a^2}{r_0^2} + o\left(\frac{a^2}{r_0^2}\right).
\eeq
In order to consider data obtained for the same value of the smoothing radius, we interpolated our data for $r_0^2 \chi^\prime_L$ as a function of $a^2 n_\cool/r_0^2 $ using a cubic spline, and considered data obtained at different values of $\beta$ and for the same value of $a^2 n_\cool /r_0^2$.

\begin{figure}[!t]
\centering
\includegraphics[scale=0.45]{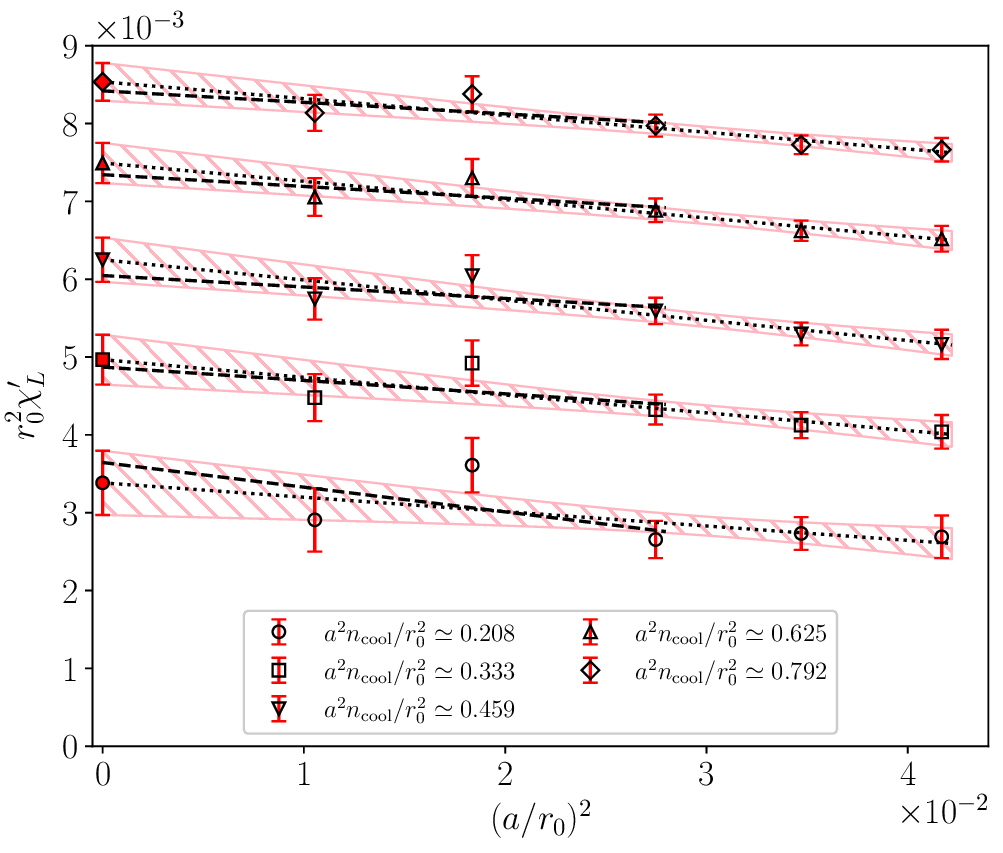}%
\caption{Examples of continuum extrapolations of $\chi^\prime_L$ for several values of the smoothing radius, expressed in physical units in terms of $a^2 n_\cool / r_0^2$. Dashed and dotted lines represent linear best fits in $(a/r_0)^2$ considering, respectively, the 3 finest lattice spacings and all available points. Shaded areas represent the fit errors on the 5-point fits.}
\label{fig:cont_limit_chip}
\end{figure}

Examples of continuum extrapolations are shown in Fig.~\ref{fig:cont_limit_chip} for several values of $a^2 n_\cool /r_0^2$. As it can be appreciated, our results for $\chi_L^\prime$ at fixed smoothing radius can be perfectly described by leading $O(a^2)$ corrections in the whole explored range of lattice spacing, and restricting our fit range to the three finest lattice spacings changes the result of the extrapolation negligibly. Thus, we take the determinations obtained from the 5-point linear fits as our final estimations of the continuum limit of $\chi^\prime(a^2 n_\cool/r_0^2)$.

As it can be observed from Fig.~\ref{fig:cont_limit_chip}, and as expected on general theoretical grounds, continuum extrapolations of $\chi^\prime$ exhibit a residual dependence on the smoothing radius. Therefore, to obtain the actual physical value of $\chi^\prime$ we need to perform a second extrapolation, this time towards the zero-cooling limit. As already discussed in Sec.~\ref{sec:lattice_topology}, we expect the continuum susceptibility slope, computed at finite smoothing radius, to exhibit a leading linear dependence on $r_s^2$. Thus, we will adopt the following function to perform the zero-smoothing-radius extrapolation:
\beq\label{eq:ansatz_fit_zerocool}
r_0^2 \chi^\prime\left(\frac{a^2 n_\cool}{r_0^2}\right) &=& r_0^2 \chi^\prime + c^\prime\, \frac{a^2 n_\cool}{r_0^2} + o\left(\frac{a^2 n_\cool}{r_0^2}\right),
\eeq
where $r_0^2 \chi^\prime$ represents our final result for the susceptibility slope.

\begin{figure}[!htb]
\centering
\includegraphics[scale=0.37]{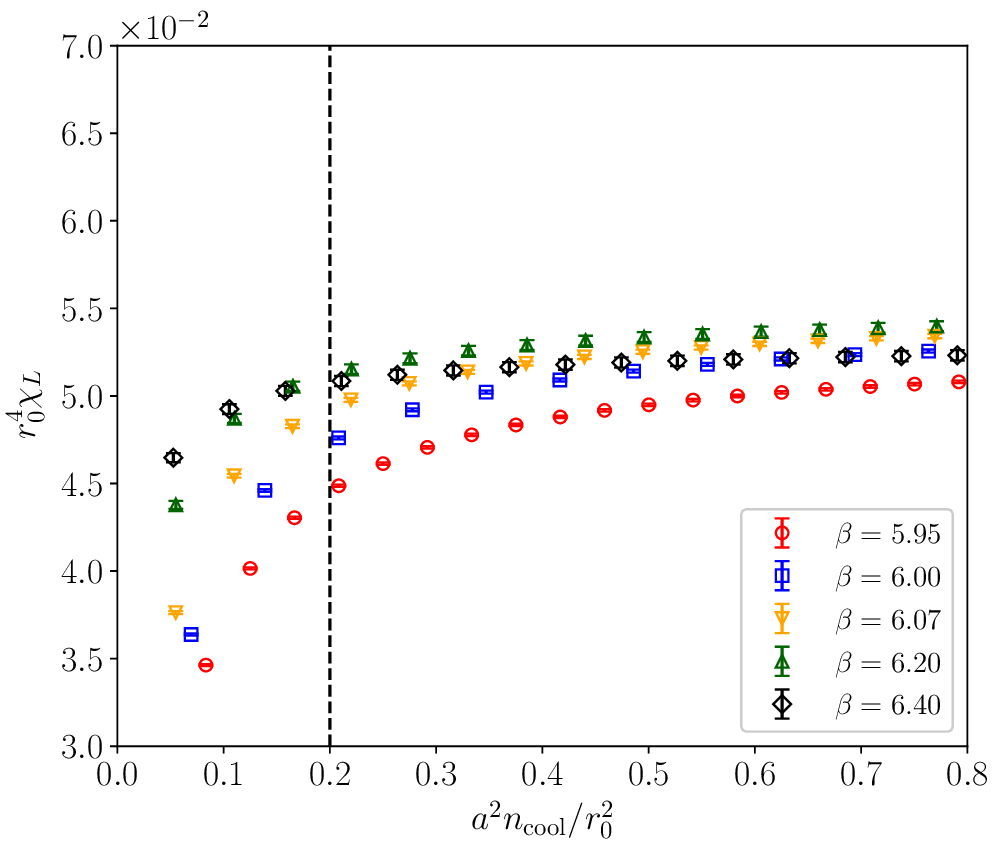}
\includegraphics[scale=0.37]{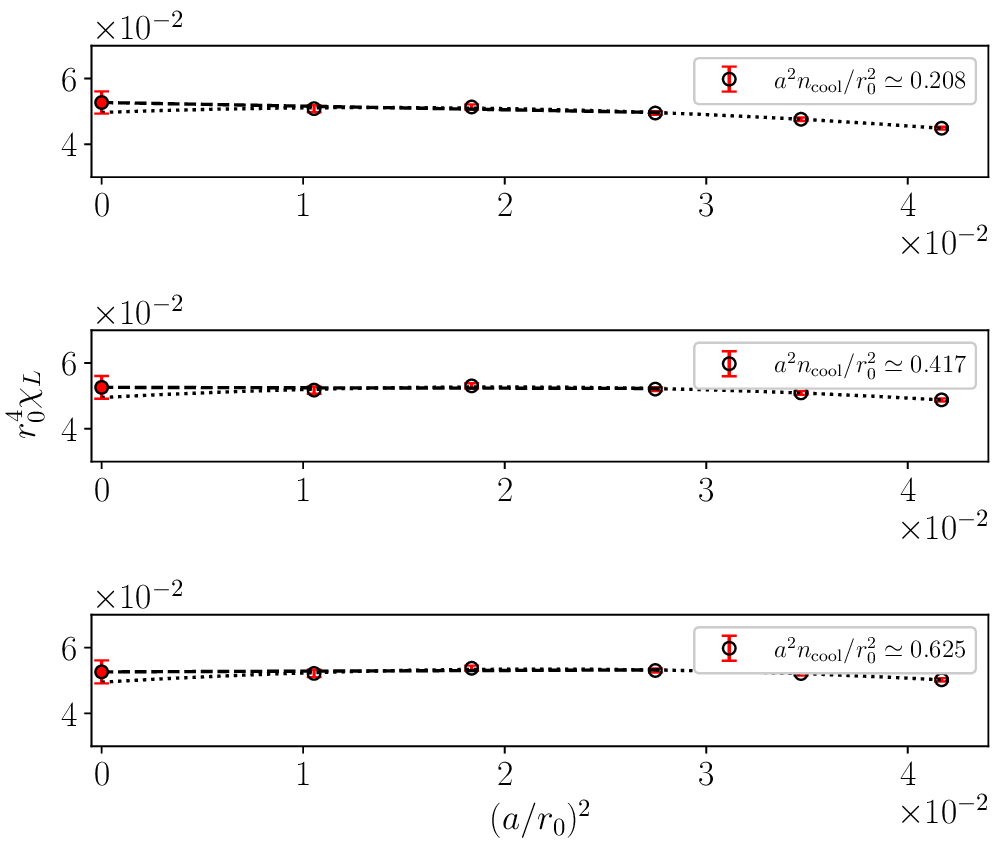}%
\includegraphics[scale=0.37]{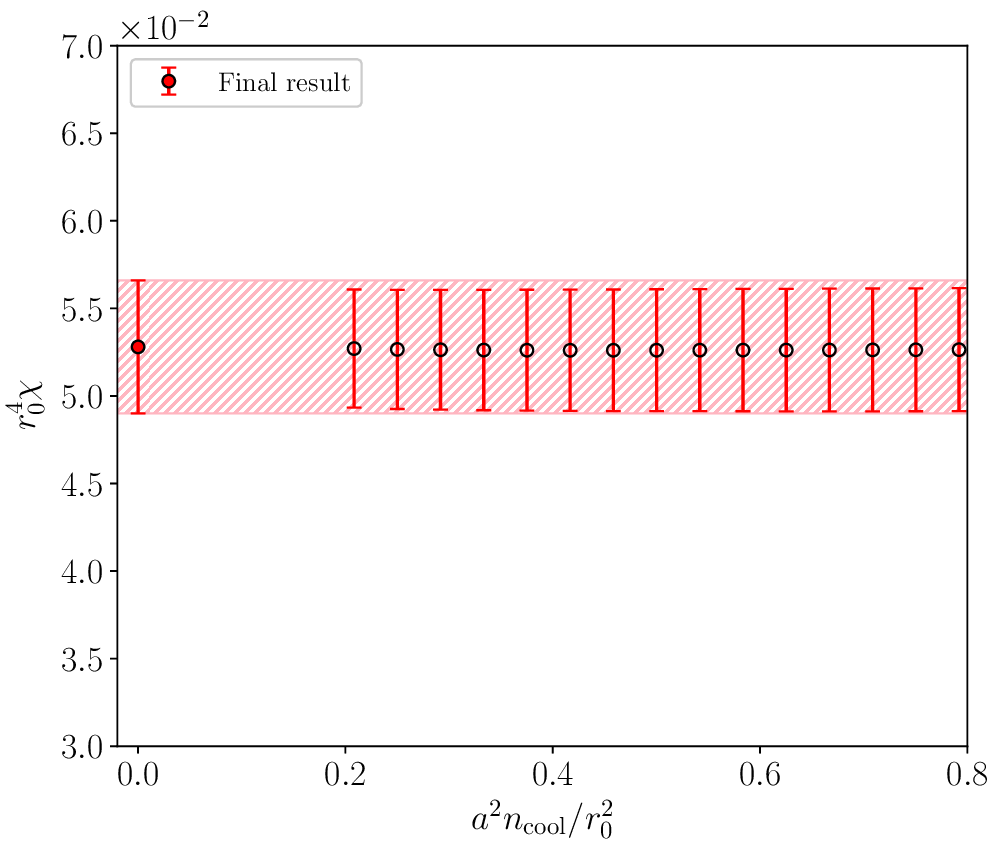}
\caption{Top panel: determinations of $r_0^4 \chi_L$ as a function of the smoothing radius, reported in terms of $a^2 n_\cool/r_0^2$, for all explored values of $\beta$. Bottom left panel: extrapolation towards the continuum limit of the lattice topological susceptibility $r_0^4 \chi_L$ at fixed smoothing radius for 3 values of $a^2 n_\cool/r_0^2$, using a linear function in $a^2$ to fit the three finest lattice spacings, and a quadratic function in $a^2$ to fit all points. Bottom right panel: obtained continuum extrapolations of $r_0^4 \chi$ as a function of $a^2 n_\cool/r_0^2$, which turn out to be fairly independent of the smoothing radius. The filled point in the origin and the related shaded area represent our final result: $r_0 \chi^{1/4} = 0.4794(86)$, i.e., $\chi^{1/4} = 200.4(3.6)~\mathrm{MeV}$ using $r_0 = 0.472(5)$ fm~\cite{Sommer:2014mea}. This result is in perfect agreement with other determinations in the literature obtained by different methods~\cite{DelDebbio:2004ns,Luscher:2010ik,Bonati:2015sqt,Athenodorou:2020ani}.}
\label{fig:cont_limit_chi}
\end{figure}

Concerning the choice of the fit range, for the lower bound we followed the same criteria adopted in Ref.~\cite{Bonanno:2022hmz}, relying on the behavior of the topological susceptibility as a function of the smoothing radius. In particular, we only considered determinations of $\chi^\prime$ obtained for smoothing radii satisfying $a^2 n_\cool / r_0^2 \ge 0.2$ (corresponding to $n_\cool \ge 5$ for the coarsest lattice spacing explored), depicted in Fig.~\ref{fig:cont_limit_chi} (top plot) as vertical dashed line. Indeed, for this range, the obtained continuum determinations of the topological susceptibility $r_0^4 \chi$ are well independent of the smoothing radius $a^2 n_\cool /r_0^2$, as shown in the bottom plots of Fig.~\ref{fig:cont_limit_chi}. The independence of $\chi$ on $r_s$ is a signal of an effective separation between the IR scale of the relevant topological fluctuations and the UV scale introduced by smoothing. This means that, staying within the plateau of $\chi$ as a function of the smoothing radius, we can reasonably expect that we did enough smoothing so as to correctly identify the topological backgrounds of our gauge configurations, but at the same time we did not do too much smoothing so as to smooth away physically-relevant topological contributions.

\begin{figure}[!htb]
\centering
\includegraphics[scale=0.55]{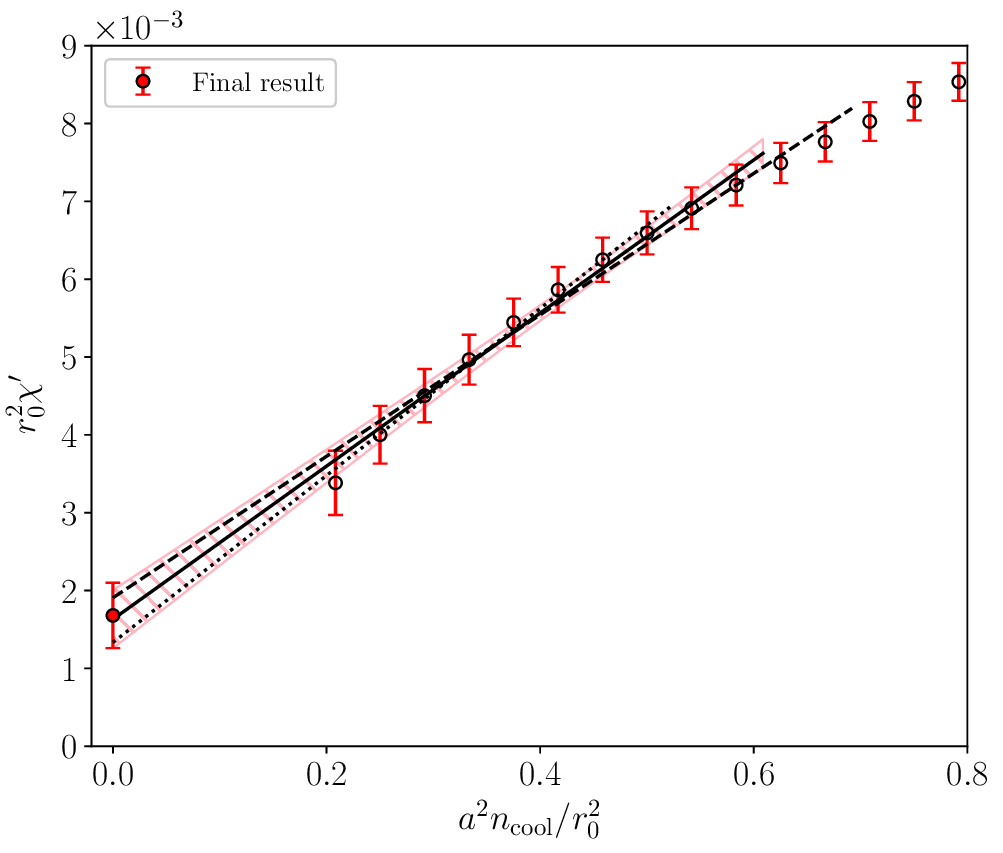}
\caption{Extrapolation of $r_0^2 \chi^\prime(a^2 n_\cool/r_0^2)$ towards the zero-smoothing-radius limit. Dotted, solid and dashed lines represent linear best fits of the results for $r_0^2 \chi^\prime(a^2 n_\cool/r_0^2)$ considering data with $a^2 n_\cool /r_0^2$ up to, respectively, 0.49, 0.58 and 0.67. The shaded area represents the fit error of the best fit performed up to $a^2 n_\cool/r_0^2 \simeq 0.58$.}
\label{fig:zerocool_limit_chip}
\end{figure}

Concerning instead the upper bound of the fit range, we observe that our data for $a^2 n_\cool / r_0^2 \lesssim 0.7$ can be well described by the linear law in Eq.~\eqref{eq:ansatz_fit_zerocool}, confirming our theoretical expectations that $\chi^\prime$ exhibits a leading linear dependence on the squared smoothing radius. We tried a few best fits varying the upper bound of the interval from $a^2 n_\cool /r_0^2 \simeq 0.5$ to $a^2 n_\cool /r_0^2 \simeq 0.7$, obtaining perfectly compatible extrapolations, cf.~Fig.~\ref{fig:zerocool_limit_chip}. In particular, we obtain $r_0^2 \chi^\prime = 0.0133(48)$, $0.00168(37)$, $0.00191(31)$ for the zero-cooling limit of the susceptibility slope when fitting our data up to, respectively, $a^2 n_\cool / r_0^2 = 0.49$, $0.58$, $0.67$.

Given the obtained values and their spread, taking $r_0^2\chi^\prime = 0.00168(37)_{\rm stat}$ for the central value and the statistical error of our final result is a perfectly reasonable choice. In order to assign a systematic error to this zero-smoothing-radius limit, we follow the prescription of Ref.~\cite{ExtendedTwistedMassCollaborationETMC:2022sta} and compute it from the difference between the final result and the zero-cooling extrapolations obtained varying the upper bound of the fit range, weighted with the probability that such difference is due to a statistical fluctuation (cf.~Eqs.~(37) and (38) in~\cite{ExtendedTwistedMassCollaborationETMC:2022sta}). We obtain $r_0^2\chi^\prime = 0.00168(37)_{\rm stat}(17)_{\rm syst}$, leading to a combined uncertainty $r_0^2 \chi^\prime = 0.00168(42)$, where the two errors were summed in quadrature.

Being the smoothing-radius dependence of $\chi^\prime$ in perfect agreement with theoretical expectations predicting a linear dependence in $n_\cool$, and given that systematics related to the linear zero-cooling extrapolation turn out to be sub-dominant compared to statistical errors, higher-order terms in $n_\cool$ appear to be unnecessary in the explored range of smoothing radii for the purpose of computing the zero-cooling limit, thus we limit ourselves to consider the results of the linear extrapolation in $a^2 n_\cool /r_0^2$. Such conclusion is perfectly in line with the smoothing-radius dependence of the time correlator of the topological charge density of the pure $\SU(3)$ gauge theory observed in Ref.~\cite{Bonanno:2023ljc}, as no higher-order term in $n_\cool$ was found to be necessary when performing the zero-cooling limit of the correlator extrapolating data obtained for $a^2 n_\cool / r_0^2 \gtrsim 0.2$ (the same lower bound considered here). In conclusion, we quote as our final result:
\beq\label{eq:final_res_chip}
\begin{aligned}
r_0^2 \chi^\prime &= 0.00168(37)_{\rm stat}(17)_{\rm syst},\\
&= 0.00168(42),\\
\\[-1em]
r_0 \sqrt{\chi^\prime} &= 0.0410(51),\\
\\[-1em]
\sqrt{\chi^\prime} &= 17.1(2.1)~\mathrm{MeV},
\end{aligned}
\eeq
where the conversion to MeV units was done using $r_0 = 0.472(5)$ fm~\cite{Sommer:2014mea}.

\subsection{Discussion of the obtained result for $\chi^\prime$}

We devote this section to a critical discussion of the obtained result for $\chi^\prime$ in Eq.~\eqref{eq:final_res_chip}.

First of all, let us recall that, according to the well-known Witten--Veneziano large-$N$ argument, the following relation holds:
\beq\label{eq:Witten_Veneziano_formula}
\widetilde{G}(0) = \chi = \frac{f_\pi^2 m^2_{\eta^\prime}}{2N_f}.
\eeq

Such relation, obtained for $p^2=0$ at large $N$, is expected to be a very good approximation even down to the physical value of the number of colors $N=3$, provided that $\widetilde{G}(p^2)$ does not change substantially from $p^2=0$ to $p^2 = m_{\eta^\prime}^2\sim O(1/N)$. This condition, using the expansion in Eq.~\eqref{eq:corr_expansion} of $\widetilde{G}(p^2)$, can be expressed through the following inequality:
\beq
\widetilde{G}(0) = \chi, \quad \widetilde{G}(p^2 = m^2_{\eta^\prime}) \simeq \chi - \chi^\prime m^2_{\eta^\prime} \qquad \implies \qquad \vert \chi^\prime \vert \, m^2_{\eta^\prime} \ll \chi.
\eeq

Using $\chi^{1/4} \simeq 200$ MeV (cf.~caption of Fig.~\ref{fig:cont_limit_chi}), $\sqrt{\chi^\prime} \simeq 17$ MeV from Eq.~\eqref{eq:final_res_chip} and $m_{\eta^\prime} \simeq 958$ MeV from the PDG~\cite{Workman:2022ynf}, we find:
\beq
\frac{\chi^\prime m_{\eta^\prime}^2}{\chi} \simeq 0.1658.
\eeq
Thus, our result for $\chi^\prime$ perfectly supports the Witten--Veneziano mechanism.

Let us now move to the comparison of our result with available phenomenological estimates for the susceptibility slope we found in the literature. The QCD Sum Rule (QCDSR) predicts\footnote{Note that the QCDSR and the ChPT results quoted in the original papers all have opposite signs compared to the numbers reported here. This is due to the fact that, in Refs.~\cite{Leutwyler:2000jg,Narison:2006ws,Narison:2021svo}, the topological charge density correlator has a global minus sign compared to our definition in Eq.~\eqref{eq:corr_def}. This leads, in particular, to $\widetilde{G}(0)=-\chi$ and $[d\widetilde{G}(p^2)/dp^2](0)=\chi^\prime$, which have opposite sign compared to, respectively, our Eqs.~\eqref{eq:chi_def} and~\eqref{eq:chiprime_def}.}~\cite{Narison:2006ws,Narison:2021svo}:
\begin{align}
\label{eq:QCDSR_puregauge}
\chi^\prime &= \textcolor{white}{+}[7(3)~\mathrm{MeV}]^2, & \quad& \text{ (QCDSR, $\SU(3)$ pure-gauge)},\\
\label{eq:chip_QCDSR_fullQCD}
\chi^\prime &= - [24.3(3.4)~\mathrm{MeV}]^2, &\quad& \text{ (QCDSR, full QCD, chiral limit)}.
\end{align}
In Chiral Perturbation Theory (ChPT) at Leading Order (LO) of the chiral expansion and assuming the presence of $N_f=3$ light flavors, one instead obtains~\cite{Leutwyler:2000jg}:
\beq\label{eq:chip_ChPT_fullQCD}
\chi^\prime = - \frac{1}{2} F_\pi^2 \left(\frac{1}{m_u^2} + \frac{1}{m_d^2} + \frac{1}{m_s^2}\right) \left(\frac{1}{m_u}+\frac{1}{m_d} + \frac{1}{m_s}\right)^{-2}.
\eeq
In the isospin-symmetric chiral limit $m_u=m_d=m_s\equiv m \to 0$, one obtains:
\beq
\chi^\prime = -\frac{1}{6} F_\pi^2 = -[ 32.8(2.4)~\mathrm{MeV} ]^2  \qquad \text{(LO ChPT, full QCD, chiral limit}),
\eeq
where we have used the latest world-average $F_\pi = 80.3(6.0)$ MeV reported in the 2021 FLAG review~\cite{FlavourLatticeAveragingGroupFLAG:2021npn} for the 3-flavor isospin-symmetric chiral limit of the pion decay constant (denoted there as $F_0$). This result is, in absolute value, larger by a factor of $\sim 1.8$ compared to the QCDSR estimate~\eqref{eq:chip_QCDSR_fullQCD}.

Looking at these determinations it is clear that the only direct comparison we are able to do is between our result and the pure-gauge QCDSR prediction~\eqref{eq:QCDSR_puregauge}. Although our number for $\chi^\prime$ is larger by a factor of $\sim 6$ compared to the QCDSR prediction, these two determinations fall in the same ballpark, and also have the same sign (which is predicted from QCDSR to change from positive to negative when going from the quenched limit to the chiral one). For the sake of completeness, we recall that also the QCDSR result for the $\SU(3)$ topological susceptibility reported in~\cite{Narison:2006ws} turns out to be a factor of $\sim 7$ smaller compared to the lattice result: $\chi(N=3)\simeq [106-122~\mathrm{MeV}]^4$ (QCDSR), $\chi(N=3)\simeq [200~\mathrm{MeV}]^4$ (lattice result).

Given the lack of other reliable predictions for the pure-gauge $\SU(3)$ susceptibility slope, we would like to conclude this discussion by presenting a ``Witten--Veneziano-like'' argument to provide an estimation of $\chi^\prime$ in the large-$N$ limit. Since there is plenty of numerical evidence that, for the pure-gauge theory, finite-$N$ corrections to the large-$N$ limit are typically small, such large-$N$ estimation should not be too far from our $N=3$ value.

First of all, let us start by summarizing very briefly how Eq.~\eqref{eq:Witten_Veneziano_formula} can be derived, as it will be useful in the following. The Euclidean topological charge density correlator in full QCD $\widetilde{G}_{\QCD}(p^2)$ can be expanded, at large-$N$, in powers of $1/N$ as:
\beq\label{eq:WV_eq}
\widetilde{G}_{\QCD}(p^2) = \widetilde{G}_{\YM}(p^2) - \frac{ \vert A_{\eta^\prime} \vert^2 }{p^2 + m^2_{\eta^\prime}} + o\left(\frac{1}{N}\right),
\eeq
where we have used that, in the large-$N$ limit, at $O(N^0)$ we have just the pure-glue (denoted with the subscript $\YM$) and the $\eta^\prime$ propagator contributions to $\widetilde{G}_{\QCD}$. The matrix element $A_{\eta^\prime} = \bra{0} q(0) \ket{\eta^\prime}$ can be estimated to be, in the large-$N$ limit:
\beq
A_{\eta^\prime} = \frac{1}{\sqrt{2N_f}} F_\pi m^2_{\eta^\prime} = \frac{1}{\sqrt{6}} F_\pi m^2_{\eta^\prime},
\eeq
where we have used $N_f=3$. Computing Eq.~\eqref{eq:WV_eq} for $p^2=0$, one obtains:
\beq\label{eq:WV2}
\chi_{\QCD} = \chi_{\YM} - \frac{1}{6}F_\pi^2 m_{\eta^\prime}^2,
\eeq
where we used the notation $\chi_{\QCD}$ and $\chi_{\YM}$ to distinguish between the full QCD and the quenched (i.e., pure-Yang--Mills) topological susceptibilities. In the chiral limit, Eq.~\eqref{eq:WV2} reduces to Eq.~\eqref{eq:Witten_Veneziano_formula}, as $\lim_{m\to 0} \chi_{\QCD}=0$:
\beq
\chi_{\YM} = \frac{1}{6}F_\pi^2 m_{\eta^\prime}^2.
\eeq

An estimate of $\chi^\prime_{\YM}$ in the large-$N$ limit can be obtained repeating exactly these steps, but considering instead $\dfrac{1}{N} \dfrac{d\widetilde{G}_{\QCD}(p^2)}{dp^2}$. The factor $1/N$ was added because, from the ChPT result in Eq.~\eqref{eq:chip_ChPT_fullQCD}, we can see that $\chi^\prime$ is $O(N)$, being $F_\pi^2 \sim O(N)$. Thus, one can expect the large-$N$ limit of $\chi^\prime/N$ to be finite and well-defined.

Deriving Eq.~\eqref{eq:WV_eq} with respect to $p^2$ and computing it in $p^2=0$, we have:
\beq
-\frac{\chi^\prime_{\QCD}}{N} = -\frac{\chi^\prime_{\YM}}{N} + \frac{1}{N}\frac{\vert A_{\eta^\prime} \vert^2}{(p^2+m_{\eta^\prime}^2)^2}\Bigg\vert_{p^2=0} = -\frac{\chi^\prime_{\YM}}{N} + \frac{1}{6}\frac{F_\pi^2}{N}.
\eeq

Taking the chiral limit, our large-$N$ prediction for $\chi^\prime_\YM/N$ finally reads:
\beq\label{eq:my_pred_chip}
\frac{\chi^\prime_{\YM}}{N} = \left[\lim_{m\to 0} \frac{\chi^\prime_{\QCD}}{N}\right] + \frac{1}{6}\frac{F_\pi^2}{N}.
\eeq

The large-$N$ limit of $F_\pi/\sqrt{N}$ has been computed in Refs.~\cite{Bali:2013kia,Perez:2020vbn} using two completely different approaches which give perfectly agreeing results. In particular, these papers report, respectively, $56(5)~\mathrm{MeV}$ and $55(5)~\mathrm{MeV}$. Concerning instead the value of $\chi^\prime_{\QCD}/N$ in the chiral limit, we can estimate it either using the ChPT prediction in Eq.~\eqref{eq:chip_ChPT_fullQCD} or the QCDSR estimation in Eq.~\eqref{eq:chip_QCDSR_fullQCD}, both obtained for $N=3$.

Putting all together, from Eq.~\eqref{eq:my_pred_chip} we obtain:
\beq
\label{eq:WVlike_chip_ChPT}
\frac{\chi^\prime_{\YM}}{N} &\simeq& (12~\mathrm{MeV})^2 \qquad \text{ (using $\chi^\prime_{\QCD}$ from ChPT)},\\
\nonumber\\[-1em]
\nonumber\\[-1em]
\label{eq:WVlike_chip_QCDSR}
\frac{\chi^\prime_{\YM}}{N} &\simeq& (18~\mathrm{MeV})^2 \qquad \text{ (using $\chi^\prime_{\QCD}$ from QCDSR)}.
\eeq
On the other hand, from our final result in Eq.~\eqref{eq:final_res_chip} we obtain:
\beq
\chi^\prime_{\YM}(N=3) = [17.1(2.1)~\mathrm{MeV}]^2 \quad \implies \quad	\frac{\chi^\prime_{\YM}(N=3)}{3} = [10.0(1.2)~\mathrm{MeV}]^2.
\eeq

As it can be observed, the sign predicted by Eqs.~\eqref{eq:WVlike_chip_ChPT} and~\eqref{eq:WVlike_chip_QCDSR} is positive in both cases, in agreement with our $N=3$ determination, and also the order of magnitude of these two large-$N$ predictions is remarkably close to our numerical $N=3$ result, especially the one in Eq.~\eqref{eq:WVlike_chip_ChPT}. Also for what concerns the internal consistency of the Witten--Veneziano mechanism we observe that the prediction in Eq.~\eqref{eq:WVlike_chip_ChPT} seems to better satisfy the condition $\vert \chi^\prime \vert \ll \chi/m^2_{\eta^\prime}$. Indeed, using $\chi/m_{\eta^\prime}^2 = F_\pi^2/6$ from Eq.~\eqref{eq:Witten_Veneziano_formula}, we find from Eq.~\eqref{eq:WVlike_chip_ChPT}: $\chi^\prime m_{\eta^\prime}^2/\chi = (\chi^\prime/N)/[F^2_\pi/(6N)] \simeq 0.2755$.

In conclusion, our ``Witten--Veneziano-like'' argument, combined with ChPT predictions, yields a large-$N$ estimation of $\chi^\prime/N$ in the quenched theory which is remarkably close to our $N=3$ lattice determination.

\section{Conclusions}\label{sec:conclu}

In this work we presented a solid lattice calculation of the topological susceptibility slope $\chi^\prime$ for the pure-gauge $\SU(3)$ Yang--Mills theory, where possible systematic effects due to finite lattice size, finite lattice spacing and finite smoothing radius are all under control.

Our numerical strategy consists in computing $\chi^\prime$ from a double limit: first, we compute the continuum limit of $\chi^\prime$ at fixed smoothing radius using results obtained for 5 different lattice spacings, then we perform a zero-cooling extrapolation using a theoretically-motivated ansatz. Concerning finite size effects, we showed, performing additional calculations for one lattice spacing varying the lattice size, that they are always negligible within our typical statistical error in the whole range of smoothing radii employed. Finally, the range of smoothing radii $r_s$ employed for the zero-cooling extrapolation was chosen so as to stay within the plateau exhibited by the continuum limit of the topological susceptibility as a function of $r_s$. As a matter of fact, we interpret the onset of such plateau as the signal of an effective separation between the IR energy scale of topological fluctuations and the UV energy scale above which fluctuations are smoothened away.

In the end, we find the following final result in physical units, obtained assuming the value $r_0 = 0.472(5)$ fm of the Sommer scale:
\beq
\chi^\prime = [17.1(2.1)~\mathrm{MeV}]^2, \qquad \text{ ($\SU(3)$, pure Yang--Mills)}.
\eeq
Our results perfectly supports the internal consistency of the Witten--Veneziano mechanism, which requires $\vert \chi^\prime \vert \ll \chi/m_{\eta^\prime}^2$. As a matter of fact, we find $\chi^\prime m_{\eta^\prime}^2 / \chi\simeq 0.166$.

Finally, we compared our results with other determinations in the literature. Concerning analytic computations, the only one that we could retrieve concerning the pure-gauge $\SU(3)$ theory is the one obtained using the QCD Sum Rule, $\sqrt{\chi^\prime}=7(3)$ MeV, which is somewhat smaller than ours, but has the same sign. We also presented a ``Witten--Veneziano-like'' argument to estimate the large-$N$ behavior of $\chi^\prime$ in the quenched theory, which assumes that $\chi^\prime$ is $\sim O(N)$ in the large-$N$ limit as predicted by Chiral Perturbation Theory. We find $\sqrt{\chi^\prime/N} \simeq 18$ MeV or $\sqrt{\chi^\prime/N} \simeq 12$ MeV , where the two estimations are obtained, respectively, using the prediction for the chiral limit of the full QCD topological susceptibility slope from the QCD Sum Rule or from Chiral Perturbation Theory. They are both remarkably close to our $N=3$ lattice result, $\sqrt{\chi^\prime(N=3)/3} = 10.0(1.2)$ MeV, especially the latter one.

In the light of the result and of the discussion here presented, it would be very interesting to pursue a direct numerical lattice investigation of the large-$N$ limit of $\chi^\prime$ in the quenched theory, adopting the same techniques here presented, in order to directly probe the $N$-dependence of $\chi^\prime$, and compare it with our prediction. We plan to investigate this interesting topic in a future publication.

\section*{Acknowledgements}
I am grateful to M.~D'Elia and M.~Garc\'ia Per\'ez for useful discussions and for reading this manuscript. This work is supported by the Spanish Research Agency (Agencia Estatal de Investigación) through the grant IFT Centro de Excelencia Severo Ochoa CEX2020- 001007-S and, partially, by grant PID2021-127526NB-I00, both funded by MCIN/AEI/ 10.13039/ 501100011033. I also acknowledge support from the project H2020-MSCAITN-2018-813942 (EuroPLEx) and the EU Horizon 2020 research and innovation programme, STRONG-2020 project, under grant agreement No 824093. Numerical calculations have been performed on the \texttt{Finisterrae~III} cluster at CESGA (Centro de Supercomputaci\'on de Galicia).

\providecommand{\href}[2]{#2}\begingroup\raggedright\endgroup

\end{document}